%% file: tarantino_hql2006.tex
\documentclass[12pt]{article}
\usepackage{epsfig}
\usepackage{amssymb,amsmath,bm}

\input econfmacros.tex

\newcommand{\bequ}{\begin{equation}}
\newcommand{\eequ}{\end{equation}}
\newcommand{\bea}{\begin{eqnarray}}
\newcommand{\eea}{\end{eqnarray}}

\newcommand{\bi}{\begin{itemize}}
\newcommand{\ei}{\end{itemize}}
\newcommand{\tev}{\, {\rm TeV}}
\newcommand{\gev}{\, {\rm GeV}}

\def\kpn{K^+\rightarrow\pi^+\nu\bar\nu}

\def\klpn{K_{L}\rightarrow\pi^0\nu\bar\nu}

\def\Title#1{\begin{center} {\Large {\bf #1} } \end{center}}

\begin{document}

\Title{Flavour Physics in the Littlest Higgs Model\\ with T-Parity}

\begin{center}{\large \bf Contribution to the proceedings of HQL06,\\
Munich, October 16th-20th 2006}\end{center}

\bigskip\bigskip

%+\addtocontents{toc}{{\it D. Reggiano}}
%+\label{ReggianoStart}

\begin{raggedright}  

{\it Cecilia Tarantino\index{Tarantino, C.}\\
Physik Department,\\
Technische Universit\"at M\"unchen,\\
D-85748 Garching, Germany}
\bigskip\bigskip
\end{raggedright}

\abstract{We present the results of an extensive analysis of flavour physics
  in both
quark and lepton sectors, in the Littlest Higgs model with T-parity (LHT).
In the quark sector, we identify some interesting scenarios for new mirror
  quark masses and $V_{Hd}$ mixing matrix that satisfy the existing
  experimental constraints from $K$ and $B$ physics and simultaneously allow
  large New Physics effects in rare decays and CP-violating observables.
In the lepton sector, where flavour violation in the
Standard Model is highly suppressed  by small neutrino masses, LHT effects turn out to
be naturally huge and could be seen in the near future measurements of lepton
flavour violating decays.}

\section{The LHT Model}

The Standard Model (SM) is in excellent agreement with the
results of particle physics experiments, in particular with
the electroweak (ew) precision measurements, thus suggesting that the SM
cutoff scale is at least as large as $10\tev$.
Having such a relatively high cutoff, however, the SM requires an
unsatisfactory fine-tuning to yield a correct ($\approx 10^2\gev$) scale for
the squared Higgs mass, 
whose corrections are quadratic and therefore highly sensitive to the cutoff.
This {\it little hierarchy problem} has been one of the main motivations to
elaborate models of physics beyond the SM.
While Supersymmetry is at present the leading candidate, 
different proposals have been formulated more recently.
Among them, Little Higgs models play an important role, being
perturbatively computable up to about $10 \tev$ and with a rather small number 
of parameters, although their predictivity can be weakened by a certain
sensitivity to the unknown ultra-violet (UV) completion of these models.

In Little Higgs models\cite{ACG} the Higgs is naturally light as it is
identified with a Nambu-Goldstone 
boson (NGB) of a spontaneously broken global
symmetry.
An exact NGB, however, would have only derivative interactions. 
Gauge and Yukawa interactions of the Higgs have to be incorporated. This can
be done without generating
quadratically divergent one-loop contributions to the Higgs mass, through the
so-called {\it collective symmetry breaking}. 
Collective symmetry breaking (SB) has the peculiarity of generating
the Higgs mass only when two or more couplings in the Lagrangian are
non-vanishing, thus avoiding one-loop quadratic divergences.
This mechanism is diagrammatically realized through the contributions of 
new particles with masses around $1 \tev$, that cancel the SM quadratic divergences.

The most economical, in matter content, Little Higgs model is the Littlest
Higgs (LH)\cite{ACKN}, where the global group $SU(5)$ is spontaneously broken
into $SO(5)$ at the scale $f \approx \mathcal{O}(1 \tev)$ and
the ew sector of the SM is embedded in an $SU(5)/SO(5)$ non-linear
sigma model. 
Gauge and Yukawa Higgs interactions are introduced by gauging the subgroup of
$SU(5)$: $[SU(2) \times U(1)]_1 \times [SU(2) \times U(1)]_2$, with gauge 
couplings respectively equal to $g_1, g_1^\prime, g_2, g_2^\prime$. 
The key feature for the realization of collective SB is that
the two gauge factors commute with a different $SU(3)$ global symmetry
subgroup of $SU(5)$, that prevents the Higgs from becoming massive when the
couplings of one of the two gauge factors vanish. 
Consequently, quadratic corrections to the squared Higgs mass involve two
couplings and cannot appear at one-loop.
In the LH model, the new particles appearing at the $\tev$ scales are the heavy
gauge bosons ($W^\pm_H, Z_H, A_H$) the heavy top ($T$) and the scalar triplet 
$\Phi$.

In the LH model, significant corrections to ew observables come
from tree-level heavy gauge boson contributions and the triplet vacuum 
expectation value (vev) which breaks the custodial $SU(2)$ symmetry. 
Consequently, ew precision tests are satisfied only for quite large
values of the New Physics (NP) scale $f \ge 2-3 \tev$\cite{HLMW,CHKMT}, unable to solve
the little hierarchy problem.
Motivated by reconciling the LH model with ew precision tests, Cheng and 
Low\cite{CL} proposed to enlarge the symmetry structure of the theory by
introducing a discrete symmetry called T-parity.
T-parity acts as an automorphism which exchanges the $[SU(2) \times U(1)]_1$ 
and $[SU(2) \times U(1)]_2$ gauge factors. The invariance of the theory under
this automorphism implies $g_1=g_2$ and $g_1^\prime = g_2^\prime$.
Furthermore, T-parity explicitly forbids the tree-level contributions of  heavy gauge bosons and the
interactions that induced the triplet vev.
The custodial $SU(2)$ symmetry is restored and the compatibility with ew
precision data is obtained already for smaller values of the NP scale, $f \ge
500 \gev$\cite{HMNP}.
Another important consequence is that particle fields are T-even or T-odd
under T-parity. The SM particles and the heavy top
$T_+$ are T-even, while the heavy gauge bosons $W_H^\pm,Z_H,A_H$ and the
scalar triplet $\Phi$ are T-odd.
Additional T-odd particles are required by T-parity: 
the odd heavy top $T_-$ and the so-called mirror fermions, i.e.,
fermions corresponding to the SM ones but with opposite T-parity and $\mathcal{O}(1 \tev)$ mass.
Mirror fermions are characterized by new flavour interactions with SM fermions
and heavy gauge bosons, which involve two new unitary 
mixing
matrices, in the quark sector, analogous to the Cabibbo-Kobayashi-Maskawa (CKM) matrix $V_{CKM}$~\cite{CKM}.
They are $V_{Hd}$ and
$V_{Hu}$, respectively involved when the SM quark is of down- or up-type,
and satisfying $V_{Hu}^\dagger V_{Hd}=V_{CKM}$\cite{HLP}.
Similarly, two new mixing matrices, $V_{H\ell}$ and $V_{H\nu}$, appear in the
lepton sector, respectively involved when the SM lepton is charged or a
neutrino and related to the PMNS matrix~\cite{pmns} through $V_{H\nu}^\dagger V_{H\ell}=V_{PMNS}^\dagger$. 
Both $V_{Hd}$ and $V_{H\ell}$
  contain $3$ angles, like $V_{CKM}$ and $V_{PMNS}$, but $3$
  (non-Majorana) phases \cite{SHORT}, i.e. 
  two additional phases relative to the SM matrices, that cannot be rotated
  away in this case.

Because of these new mixing matrices, the LHT model does not belong to the Minimal
Flavour Violation (MFV) class of models~\cite{UUT,AMGIISST} and significant 
effects in flavour observables are possible.
Other LHT peculiarities are the rather small number of new particles and
parameters (the SB scale $f$, the parameter $x_L$ describing $T_+$ mass and
interactions, the mirror fermion masses and $V_{Hd}$ and $V_{H\ell}$
parameters) and the
absence of new operators in addition to the SM ones.
On the other hand, one has to recall that Little Higgs models are low
energy non-linear sigma models, whose unknown UV-completion introduces a
theoretical uncertainty reflected by a left-over logarithmic cut-off dependence~\cite{BPUB,BBPRTUW} in $\Delta F=1$ processes.

\section{LHT Flavour Analysis}

Several studies of flavour physics in the LH model without T-parity have been
performed in the last four years \cite{BPUB,FlavLH}. Without T-parity, mirror 
fermions and new sources of flavour and CP-violation are absent, the LH model
is a MFV model and NP contributions result to be very small. 

More recently, flavour physics analyses have been also performed in the LHT
model, for both quark~\cite{HLP,BBPRTUW,BBPTUW} and lepton sectors~\cite{Indian,BBDPT}.
In this model, new mirror fermion interactions can yield large NP effects,
mainly in $K$ and $B$ rare and CP-violating decays and in lepton flavour violating decays. 

\subsection{LHT Analysis in the Quark Sector}

In~\cite{BBPRTUW,BBPTUW} we have studied in the LHT model $B$ and $K$ meson mixings,
CP-violation, rare decays and the radiative decay $B \rightarrow X_s \gamma$.
We have imposed well known experimental constraints and estimated LHT effects
in those observables that are not yet measured or still very uncertain.
We have considered several scenarios for the
structure of the $V_{Hd}$ matrix and the mass spectrum of mirror
quarks in order to gain a global view over
possible LHT signatures.
The parameters $f$ and $x_L$ have been
fixed to $f=1
\tev$ and $x_L=0.5$ in accordance with ew precision tests~\cite{HMNP}.
The CKM parameters entering the analysis have been taken from tree
level decays only, where NP effects can be neglected.
In order to simplify the numerical analysis we have set all non-perturbative 
parameters to their central values, while allowing $\Delta M_K$, 
$\varepsilon_K$, $\Delta M_d$, $\Delta M_s$, $\Delta M_s / \Delta M_d$ and $S_{\psi K_S}$ to differ from 
their experimental values by $\pm 50\%$, $\pm 40\%$, $\pm 40\%$, $\pm 40\%$,
$\pm 20\%$ and $\pm 8\%$, respectively. This rather conservative choice
guarantees that important effects are not missed.

Two interesting scenarios have been identified.
In the first one (B-scenario) large enhancements in $B$ physics
are possible, while in the second one (K-scenario) important effects appear in
$K$ observables.
They are both characterized by the quasi-degeneracy of the first two mirror
quark generations ($m_{H1} \simeq m_{H2} \simeq 500 \gev$, $m_{H3} \simeq 1000
\gev$), as required by $\Delta M_K$ and $\varepsilon_K$ constraints.
The new mixing angles in $V_{Hd}$ are chosen to satisfy the hierarchy
$s^d_{23}\ll s^d_{13}\le s^d_{12}$ in B-scenario
and the hierarchy $s^d_{23}\simeq s^d_{13}< s^d_{12}=1/\sqrt{2}$ in K-scenario.
Moreover, the two additional phases of $V_{Hd}$, whose impact is
numerically small, have been set to zero.
In addition, in order to explore all possible LHT effects, we have
performed a general
scan over mirror quark masses and $V_{Hd}$ parameters.
In the following scatter plots, B- and K-scenarios and general scan are
respectively displayed as green, brown and blue points, while red points
correspond to a less general scan over $V_{Hd}$ parameters at fixed mirror
masses ($m_{H1} = 400 \gev$, $m_{H2} = 500 \gev$, $m_{H3} = 600 \gev$).
\begin{figure}
\center{\epsfig{file=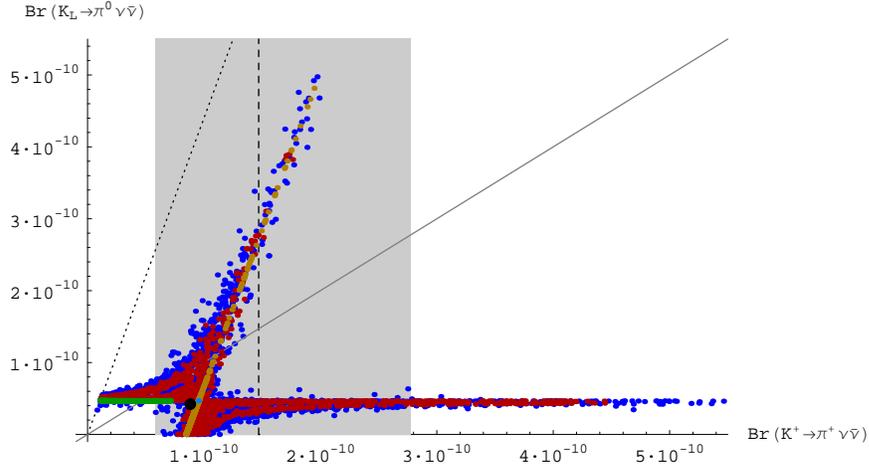,scale=0.7}}
\caption{\it $Br(\klpn)$ as a function of $Br(\kpn)$. The shaded
    area represents the experimental $1\sigma$-range for $Br(\kpn)$. The
    model-independent Grossman-Nir bound~\cite{GNbound} is displayed by the dotted line, while the solid line
    separates the two areas where $Br(\klpn)$ is larger or smaller than
    $Br(\kpn)$.}
\label{fig:KLKp}
\end{figure}

\begin{figure}
\center{\epsfig{file=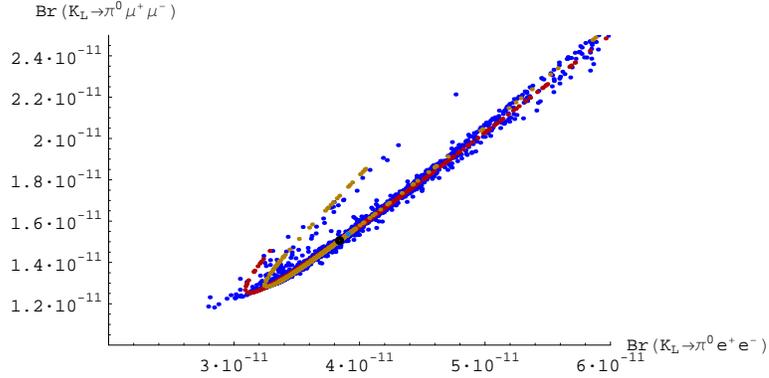,scale=0.7}}
\caption{\it $Br(K_L \to \pi^0 \mu^+\mu^-)$ as a
  function of $Br(K_L\to \pi^0 e^+e^-)$.}
\label{fig:KmuKe}
\end{figure}

The main results of our LHT analysis~\cite{BBPRTUW,BBPTUW} in the quark sector
are:
\begin{itemize}

\item {The most evident departures from the SM predictions are found for
    CP-violating observables that are strongly suppressed in the SM. 
These are the branching ratio for $K_L \to \pi^0 \nu \bar \nu$
    (Fig.~\ref{fig:KLKp}) and
    the CP-asymmetry $S_{\psi \phi}$, that can be enhanced by an order of
    magnitude relative to the SM predictions. Large departures from SM expectations are also possible for $Br(K_L \to
  \pi^0 \ell^+ \ell^-)$ (Fig.~\ref{fig:KmuKe}), $Br(K^+ \to \pi^+ \nu \bar \nu)$
    (Fig.~\ref{fig:KLKp}) and the
    semileptonic CP-asymmetry $A^s_\text{SL}$, that can be enhanced by an
    order of magnitude w.r.t the SM.}

\item {The branching ratios for $B_{s,d} \to \mu^+ \mu^-$ and $B \to X_{s,d}
    \nu \bar \nu$, instead,  are modified by at most $50\%$ and $35\%$,
    respectively, and the effects of new electroweak penguins in $B \to \pi K$
    are small, in agreement with the recent data.
The new physics effects in $B\to X_{s,d}\gamma$ and $B\to X_{s,d}\ell^+\ell^-$
turn out to be below $5\%$ and $15\%$, respectively, so that agreement 
with the data can easily be obtained.}

\item {Small, but still significant effects have been found in $B_{s,d}$ mass
    differences. In particular, a $7\%$ suppression of $\Delta M_s$ is
    possible, thus improving the compatibility with the recent experimental
    measurement~\cite{CDFD0}.}

\item {The
possible ``discrepancy''~\cite{UTfit,CKMfit,BBGT} between the values of $\sin 2\beta $
following directly from $A_{\rm CP}(B_d \to \psi K_S)$ and
indirectly from the analysis of the unitarity triangle
involving only tree-level processes, and in particular $|V_{ub}|$, can
be cured within the LHT model thanks to a new phase $\varphi_{B_d}\simeq -5^o$.}
\item{The universality of new physics effects, characteristic for MFV models,
    can be largely broken, in particular between $K$ and $B_{s,d}$ systems.
NP effects, in fact, are typically larger in $K$ system where the SM contribution is
    CKM-suppressed.
In particular, sizable departures from MFV relations between $\Delta M_{s,d}$
    and $Br(B_{s,d} \to \mu^+ \mu^-)$ and between $S_{\psi K_S}$ and the $K \to
    \pi \nu \bar \nu$ decay rates are possible.}

\end{itemize}

\subsection{LHT Analysis in the Lepton Sector}
\begin{table}[!]
\begin{center}
\begin{tabular}{|c|c|c|c|}
\hline
decay & $f=1000\gev$ & $f=500\gev$ & exp.~upper bound \\\hline\hline
$\mu\to e\gamma$ & $1.2\cdot 10^{-11}$ ($1\cdot10^{-11}$) & $1.2\cdot 10^{-11}$ ($1\cdot 10^{-11}$) & $1.2\cdot 10^{-11}$ \cite{muegamma} \\
$\mu^-\to e^-e^+e^-$ & ~$1.0\cdot 10^{-12}$ ($1\cdot 10^{-12}$)~ &~$1.0\cdot 10^{-12}$ ($1\cdot 10^{-12}$)~ & $1.0\cdot 10^{-12}$ \cite{meee} \\
$\mu\text{Ti}\to e\text{Ti}$  & $2\cdot 10^{-10}$ ($5\cdot10^{-12}$) & $4\cdot10^{-11}$ ($5\cdot10^{-12}$) & $4.3\cdot10^{-12}$ \cite{mue-conv_bound}  \\\hline
$\tau\to e\gamma$ & $8\cdot 10^{-10}$ ($7\cdot 10^{-10}$) & ${1\cdot 10^{-8}}$ (${1\cdot 10^{-8}}$) & ${9.4\cdot10^{-8}}$ \cite{Banerjee} \\
$\tau\to \mu\gamma$ & $8\cdot 10^{-10}$ ($8\cdot 10^{-10}$) &$2\cdot 10^{-8}$  (${1\cdot 10^{-8}}$) &${1.6\cdot10^{-8}}$ \cite{Banerjee}\\
$\tau^-\to e^-e^+e^-$ & $7\cdot10^{-10}$ ($6\cdot 10^{-10}$) & ${2\cdot10^{-8}}$  (${2\cdot 10^{-8}}$) & $2.0\cdot10^{-7}$ \cite{AUBERT04J}\\
$\tau^-\to \mu^-\mu^+\mu^-$ & $7\cdot10^{-10}$ ($6\cdot 10^{-10}$) & ${3\cdot10^{-8}}$  (${3\cdot 10^{-8}}$)  & $1.9\cdot10^{-7}$ \cite{AUBERT04J} \\
$\tau^-\to e^-\mu^+\mu^-$ & $5\cdot10^{-10}$ ($5\cdot 10^{-10}$)& ${2\cdot10^{-8}}$ (${2\cdot 10^{-8}}$)  & $2.0\cdot10^{-7}$ \cite{YUSA04}\\
$\tau^-\to \mu^-e^+e^-$ & $5\cdot10^{-10}$ ($5\cdot 10^{-10}$)& ${2\cdot10^{-8}}$ (${2\cdot 10^{-8}}$) &$1.9\cdot10^{-7}$ \cite{YUSA04} \\
$\tau^-\to \mu^-e^+\mu^-$ & $5\cdot10^{-14}$ ($3\cdot10^{-14}$) & ${2\cdot10^{-14}}$ (${2\cdot10^{-14}}$) & $1.3\cdot10^{-7}$ \cite{AUBERT04J}\\
$\tau^-\to e^-\mu^+e^-$ & $5\cdot10^{-14}$ ($3\cdot10^{-14}$) &${2\cdot10^{-14}}$ (${2\cdot10^{-14}}$)  & $1.1\cdot10^{-7}$ \cite{AUBERT04J} \\
$\tau\to\mu\pi$ & $2\cdot10^{-9} $ ($2\cdot10^{-9} $) & ${5.8\cdot10^{-8}}$ (${5.8\cdot10^{-8}}$) & ${5.8\cdot10^{-8}}$ \cite{Banerjee}\\
$\tau\to e\pi$ & $2\cdot10^{-9} $ ($2\cdot10^{-9} $)& ${4.4\cdot10^{-8}}$ (${4.4\cdot10^{-8}}$)   & ${4.4\cdot10^{-8}}$ \cite{Banerjee}\\
$\tau\to\mu\eta$ & $6\cdot10^{-10}$ $(6\cdot10^{-10})$ & ${2\cdot10^{-8}}$ ${(2\cdot10^{-8})}$ &  ${5.1\cdot 10^{-8}}$ \cite{Banerjee}\\
$\tau\to e\eta$ & $6\cdot10^{-10}$ $( 6\cdot10^{-10})$ & ${2\cdot10^{-8}}$ ${(2\cdot10^{-8})}$ &  ${4.5\cdot 10^{-8}}$ \cite{Banerjee}\\
$\tau\to \mu\eta'$ & $7\cdot10^{-10}$ $(7\cdot10^{-10})$& ${3\cdot10^{-8}}$ ${(3\cdot10^{-8})}$ & ${5.3\cdot 10^{-8}}$ \cite{Banerjee}\\
$\tau\to e\eta'$ & $7\cdot10^{-10}$ $(7\cdot10^{-10})$& ${3\cdot10^{-8}}$ ${(3\cdot10^{-8})}$  & ${9.0\cdot 10^{-8}}$ \cite{Banerjee}\\\hline
$K_L\to\mu e$ & $4\cdot 10^{-13}$ ($2\cdot10^{-13}$) &  $3\cdot10^{-14}$ ($3\cdot10^{-14}$)& $4.7\cdot10^{-12}$ \cite{KLmue-exp}\\
$K_L\to\pi^0\mu e$ & $4\cdot 10^{-15}$ ($2\cdot10^{-15}$)  &  $5\cdot10^{-16}$ ($5\cdot10^{-16}$) & $6.2\cdot10^{-9}$ \cite{ARISAKA98}\\
$B_d\to\mu e$ & $5\cdot10^{-16}$ ($2\cdot10^{-16}$)  & $9\cdot10^{-17}$ ($9\cdot10^{-17}$) & $1.7\cdot10^{-7}$ \cite{CHANG03}\\
$B_s\to\mu e$ & $5\cdot 10^{-15}$ ($2\cdot10^{-15}$)  &$9\cdot10^{-16}$ ($9\cdot10^{-16}$) & $6.1\cdot10^{-6}$ \cite{ABE98V}\\
$B_d\to\tau e$ & $3\cdot 10^{-11}$  ($2\cdot10^{-11}$) & ${2\cdot10^{-10}}$ ($2\cdot10^{-10}$) & $1.1\cdot10^{-4}$ \cite{BORNHEIM04}\\
$B_s\to\tau e$ & $2\cdot10^{-10}$ ($2\cdot10^{-10}$) & ${2\cdot10^{-9}}$ ($2\cdot10^{-9}$)& ---\\
$B_d\to\tau\mu$ & $3\cdot10^{-11}$ ($3\cdot10^{-11}$) & $3\cdot10^{-10}$ ($3\cdot10^{-10}$) & $3.8\cdot10^{-5}$ \cite{BORNHEIM04} \\
$B_s\to\tau\mu$ & $2\cdot10^{-10}$ ($2\cdot10^{-10}$) &  $3\cdot10^{-9}$ ($3\cdot10^{-9}$)& ---\\\hline
\end{tabular}
\end{center}
\caption{\it Upper bounds on LFV decay branching ratios in the LHT model, for two different values of the scale $f$, after imposing the constraints on $\mu\to e\gamma$ and $\mu^-\to e^-e^+e^-$. The numbers given in brackets are obtained after imposing the additional constraint $R(\mu\text{Ti}\to e\text{Ti})<5\cdot10^{-12}$. {For $f=500\gev$, also the bounds on $\tau\to\mu\pi,e\pi$ have been included.} The current experimental upper bounds are also given.\label{tab:bounds}}
\end{table}

\begin{figure}
\center{\epsfig{file=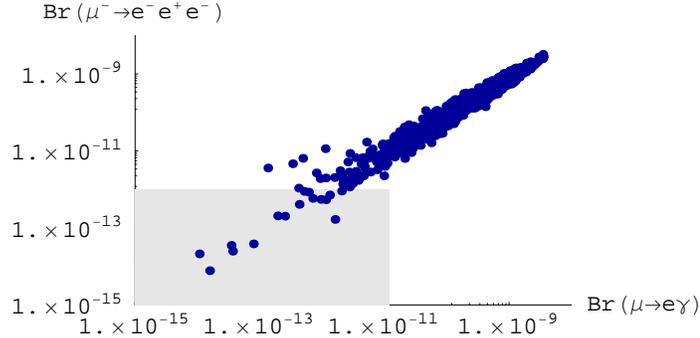,scale=0.9}}
\caption{\it Correlation between the branching ratios for $\mu\to e\gamma$ and
  $\mu^-\to e^-e^+e^-$ from a general scan over the LHT parameters. The shaded
  area represents present experimental upper bounds.}
\label{fig:mue}
\end{figure}
In contrast to rare K and B decays, where the SM contributions play an
important and often dominant role in the LHT model, the smallness of ordinary
neutrino masses assures that mirror fermion contributions to lepton flavour
violating (LFV) processes are by far the dominant effects.
Moreover, the absence of QCD corrections and hadronic matrix elements allows
in most cases to make predictions entirely within perturbation theory.

In~\cite{BBDPT} we have studied the most interesting LFV processes: $\ell_i
\rightarrow \ell_j \gamma$, $\tau \rightarrow \ell P$ (with $P=\pi, \eta, \eta'$), $\mu^- \rightarrow e^-
e^+ e^-$, the six three-body decays $\tau^- \rightarrow l_i^- l_j^+ l_k^-$ and 
the rate for $\mu-e$ conversion in nuclei.
We have also calculated the rates for $K_{L,S} \rightarrow \mu e$, $K_{L,S}
\rightarrow \pi^0 \mu e$, $B_{d,s} \rightarrow \mu e$, $B_{d,s} \rightarrow
\tau e$ and $B_{d,s} \rightarrow \tau \mu$.

At variance with meson decays, the number of flavour violating decays in the
lepton sector, for which significant experimental constraints exist, is rather
limited.
Basically only the upper bounds on $Br(\mu \rightarrow e \gamma)$, $Br(\mu^-
\rightarrow e^- e^+ e^-)$, $Br(K_L \rightarrow \mu e)$ and $R(\mu \text{Ti}
\to e \text{Ti})$ can be used in our analysis.
The situation may change significantly in the coming years thanks to near
future experiments~\cite{mue-conv_bound,megexp,SuperB,J-PARK}.
Meanwhile, we have estimated LHT effects, imposing the experimental bounds 
mentioned above and scanning over mirror lepton masses in the range $[300
\gev, 1500 \gev]$ and over the parameters of the $V_{H\ell}$ mixing matrix,
with the symmetry breaking scale $f$ fixed to $f=1 \tev$ or $f=500 \gev$ 
in accordance with ew precision tests~\cite{HMNP}.
We note that for $f=500 \gev$ also the very recent experimental upper bounds on
$\tau \to \mu \pi, e \pi$ given in ~\cite{Banerjee}, where
Belle~\cite{Belle-radiative,Belle-semi} and BaBar~\cite{teg-exp,BaBar-semi} results have been combined, become effective.

We have found that essentially all the rates considered can reach or approach
present experimental upper bounds, as shown in table~\ref{tab:bounds}.
In particular, in order to suppress the $\mu \rightarrow e \gamma$ and $\mu^-
\rightarrow e^- e^+ e^-$ decay rates below the experimental upper bounds (see Fig.~\ref{fig:mue}), the
$V_{H\ell}$ mixing matrix has to be rather hierarchical, unless mirror
leptons are quasi-degenerate. 

Moreover, following the strategy proposed
in~\cite{Ellis:2002fe,Arganda:2005ji,Paradisi1} in the supersymmetric framework, we have identified certain correlations between
branching ratios that are less parameter dependent than the individual branching ratios
and could provide a clear signature of the model.
In particular, we find that the ratios $Br(\ell_i \to \ell_j \ell_j
\ell_j)/Br(\ell_i \to \ell_j \gamma)$, $Br(\ell_i \to \ell_j \ell_j
\ell_j)/Br(\ell_i \to \ell_j \ell_k \ell_k)$ and $Br(\ell_i \to \ell_j \ell_k
\ell_k)/Br(\ell_i \to \ell_j \gamma)$ could allow for a transparent
distinction between the LHT model and the MSSM (see Table~\ref{tab:comparison}).
\begin{table}
{\renewcommand{\arraystretch}{1.5}
\begin{center}
\begin{tabular}{|c|c|c|c|}
\hline
ratio & LHT  & MSSM (dipole) & MSSM (Higgs) \\\hline\hline
$\frac{Br(\mu^-\to e^-e^+e^-)}{Br(\mu\to e\gamma)}$  & \hspace{.8cm} 0.4\dots2.5\hspace{.8cm}  & $\sim6\cdot10^{-3}$ &$\sim6\cdot10^{-3}$  \\
$\frac{Br(\tau^-\to e^-e^+e^-)}{Br(\tau\to e\gamma)}$   & 0.4\dots2.3     &$\sim1\cdot10^{-2}$ & ${\sim1\cdot10^{-2}}$\\
$\frac{Br(\tau^-\to \mu^-\mu^+\mu^-)}{Br(\tau\to \mu\gamma)}$  &0.4\dots2.3     &$\sim2\cdot10^{-3}$ & $0.06\dots0.1$ \\
$\frac{Br(\tau^-\to e^-\mu^+\mu^-)}{Br(\tau\to e\gamma)}$  & 0.3\dots1.6     &$\sim2\cdot10^{-3}$ & $0.02\dots0.04$ \\
$\frac{Br(\tau^-\to \mu^-e^+e^-)}{Br(\tau\to \mu\gamma)}$  & 0.3\dots1.6    &$\sim1\cdot10^{-2}$ & ${\sim1\cdot10^{-2}}$\\
$\frac{Br(\tau^-\to e^-e^+e^-)}{Br(\tau^-\to e^-\mu^+\mu^-)}$     & 1.3\dots1.7   &$\sim5$ & 0.3\dots0.5\\
$\frac{Br(\tau^-\to \mu^-\mu^+\mu^-)}{Br(\tau^-\to \mu^-e^+e^-)}$   & 1.2\dots1.6    &$\sim0.2$ & 5\dots10 \\
$\frac{R(\mu\text{Ti}\to e\text{Ti})}{Br(\mu\to e\gamma)}$  & $10^{-2}\dots 10^2$     & $\sim 5\cdot 10^{-3}$ & $0.08\dots0.15$ \\\hline
\end{tabular}
\end{center}\renewcommand{\arraystretch}{1.0}
}
\caption{\it Comparison of various ratios of branching ratios in the LHT model and in the MSSM without and with significant Higgs contributions.\label{tab:ratios}}
\end{table}

Finally, we have studied the muon anomalous magnetic moment finding that,
even for values of the NP scale $f$ as low as $500 \gev$, $a_\mu^\text{LHT}<1.2\cdot 10^{-10}$.
This value is roughly a factor $5$ below the current experimental
uncertainty~\cite{Bennett:2006fi}, implying that the possible discrepancy between the SM prediction
and the data cannot be solved in the LHT model.

\bigskip
I would like to thank the organizers of the interesting and
pleasant conference {\it Heavy Quarks and Leptons} realized in Munich.
Special thanks go to the other authors of the work presented here:
Monika Blanke, Andrzej J. Buras, Bj\"orn Duling, Anton Poschenrieder, Stefan
Recksiegel, Selma Uhlig and Andreas Weiler.

\end{document}

%% file: econfmacros.tex
\def\beq{\begin{equation}}
\def\eeq#1{\label{#1}\end{equation}}
\def\eeqn{\end{equation}}

\def\beqa{\begin{eqnarray}}
\def\eeqa#1{\label{#1}\end{eqnarray}}
\def\eeqan{\end{eqnarray}}

\let\bar=\overbar

\def\Dslash{\not{\hbox{\kern-4pt $D$}}}
\def\dslash{\not{\hbox{\kern-2pt $\del$}}}

\def\msb{{\bar{\ssstyle M \kern -1pt S}}}